\documentclass[amssymb,aps,twocolumn,floats,showpacs,prb]{revtex4-1}
\usepackage{color,graphicx}

\newcommand{\YK}{120$^{\circ}~$}

\begin{document}
\title{Noncollinear and noncoplanar magnetic order in the extended Hubbard model on
anisotropic triangular lattice}

\author{Kanika Pasrija and Sanjeev Kumar}
\affiliation{Indian Institute of Science Education and Research(IISER) Mohali,
Sector 81, S.A.S. Nagar, Manauli PO 140 306, India}

\date{\today}


\begin{abstract}
Motivated by the importance of non-collinear and non-coplanar magnetic phases in determining various electrical
properties of magnetic materials, we investigate the phase diagrams
of the extended Hubbard model on anisotropic triangular lattice. We make use of a mean-field scheme that 
treats collinear, non-collinear and non-coplanar phases on equal footing. 
In addition to the ferromagnetic and \YK antiferromagnetic phases, we find the four-sublattice flux, the 3Q non-coplanar
and the non-collinear charge-ordered states to be stable at specific values of filling fraction $n$.
Inter-site Coulomb repulsion leads to intriguing spin-charge ordered phases. Most notable of these are
the collinear and non-collinear magnetic states at $n=2/3$, which occur together with 
a pinball-liquid-like charge order. 
Our results demonstrate that the elementary single-orbital extended Hubbard model on a triangular lattice hosts 
unconventional spin-charge ordered phases, which have been reported in more complex and material-specific electronic Hamiltonians
relevant to layered triangular systems.
\end{abstract}

\maketitle

\section{Introduction}

The transition metals and their oxides are well known for exhibiting a 
variety of magnetic ordering phenomena \cite{Zener1951,Maekawa2004}. The nature of the low-temperature
magnetically ordered states depends on the type of dominant magnetic exchange interactions, which
in turn depend on the details of the crystal structure and electronic band structure. 
The theoretical models that describe magnetism in solids can be divided into three broad categories: 
(i) models that do not {\it apriori} contain a local moment but allow for moment formation via electron-electron interactions, 
such as the Hubbard model \cite{Hubbard1963}, (ii) models consisting of localized spins only, such as the Heisenberg model,
and (iii) models that have itinerant electrons coupled to localized moments, e.g., the Kondo lattice model \cite{Stewart1984}. 
Starting with a general multi-orbital Hubbard model, the other two types can be obtained in appropriate limits \cite{Auerbach2012}. Hence, the Hubbard model
is the elementary model for describing magnetism. Indeed, two of the most common magnetically ordered ground states, the ferromagnet and the staggered antiferromagnet,
are present in the mean-field phase diagram of the Hubbard model \cite{Hirsch1985}.

Search for non-collinear and non-coplanar magnetic ordering in models and materials has emerged as 
an important research topic in recent years. The reason being the fundamental connection between the nature of magnetic order and electrical properties of various magnetic materials.
A non-coplanar magnetic order is known to give rise to 
anomalous Hall response in transport \cite{Nagaosa2010,Kezsmarki2005,Hayami2015}. On the other hand, a planar spiral order can allow for 
a ferroelectric response via the spin-current mechanism \cite{Katsura2005,Mostovoy2006}. 
Recent theoretical studies on Kondo-lattice models have shown that a variety of unconventional magnetic ground states can be
stabilized depending on the underlying lattice geometry and electronic filling fraction. In particular, triangular and checkerboard lattices allow for non-coplanar
magnetism at quarter filling of the band \cite{Martin2008,Kumar2010,Akagi2010a,Akagi2012,Ishizuka2012,Ishizuka2012a,Venderbos2012}. 
More recently, existence of spin-charge ordered phases has also been reported at the average filling fraction of two electrons per three sites \cite{Reja2015}.
Given that the Hubbard model is the fundamental model for magnetism, it is important to know if the unconventional magnetic phases found in the Kondo-lattice
Hamiltonian are also present in the Hubbard model. Moreover, for some of the magnetic phases the shape of the Fermi surface plays a crucial role
and the ordered phases should therefore be independent of the nature of interactions \cite{Martin2008, Chern2010}. 
To the best of our knowledge a systematic search across the electronic density range for non-collinear and non-coplanar magnetic phases in the Hubbard model on triangular lattice has not been reported.

In this work, we map out the magnetic phase diagram of the Hubbard model on anisotropic triangular lattice within a variational mean-field approach that 
captures the non-collinear and non-coplanar magnetism on same footing as the collinear magnetism. The phase diagram is rich and consists of many unconventional phases that are found in 
the corresponding Kondo-lattice model. 
These include, (i) the four-sublattice 3Q order at quarter filling, (ii) the flux state at quarter filling, 
(iii) the non-collinear charge ordered phase at two-third filling. In addition, the \YK state is stable near the isotropic triangular limit and
a collinear stripe state is stable in the anisotropic regime. The influence of nearest neighbor Coulomb repulsion 
on these phases is also investigated. We find that a pinball-liquid like charge ordering phase, 
which has so far been reported in spinless fermion models or multi-orbital models on triangular lattice \cite{Hotta2006,Hotta2006a,Miyazaki2009,Ralko2015}, 
is stabilized by the Coulomb interactions at a filling fraction of $n=2/3$ and occurs concomitantly with either collinear or non-collinear magnetic order. 
Given the fundamental nature of the Hubbard Hamiltonian, our results are of general interest. In addition, the reported phases are of specific 
relevance to various triangular lattice systems, such as the layered AgNiO$_2$ \cite{Wawrzynska2008,Wawrzynska2007a}, 
the NaCoO$_2$ \cite{Mukhamedshin2014,Mukhamedshin2014a,Mukhamedshin2004,Bernhard2004,Alloul2009} and the organic charge-transfer salts \cite{Hotta2012,Kagawa2013}.

The remainder of the paper is organized as follows. In section II we describe
the model and briefly discuss the previous investigations that focus on non-collinear magnetism. In section III we describe the
method used in this work. Results and discussions follow in
Section IV, where we first discuss the density versus interaction phase diagram,
and then discuss the phase diagrams corresponding to specific electronic filling fractions.
In section V we analyze the effect of nearest neighbor Coulomb repulsion on the magnetic 
phase diagram. Conclusions are presented in section VI.

\section{Model}
The single-orbital Hubbard Hamiltonian on an anisotropic triangular lattice is given by,

\begin{eqnarray}
H &=& -t(t') \sum_{\langle ij \rangle, \sigma}
c^{\dagger}_{i\sigma}c^{}_{j\sigma} - \mu \sum_i n_i 
+  U \sum_{i} n_{i\uparrow} n_{i\downarrow},
\end{eqnarray}
where $c^{\dagger}_{i\sigma}(c^{}_{i\sigma})$ is the creation (annihilation)
operator, and 
$n_{i\sigma} = c^{\dagger}_{i\sigma}c^{}_{i\sigma}$ is the number operator for
electrons with spin $\sigma = \uparrow, \downarrow$ at site $i$. 
The total number operator for electrons at site $i$ is $n_i = n_{i \uparrow} +
n_{i \downarrow}$, and $\mu$ is the chemical potential.
The electronic hopping amplitudes between 
nearest neighbor sites on
triangular lattice are $t$ and $t'$ as shown in Fig. 1, and $U$ denotes the
on-site Hubbard repulsion.

The Hubbard model is one of the most successful and well studied Hamiltonians in condensed matter physics \cite{LeBlanc2015}. However, due to its 
relevance to various materials, certain filling fractions have been of more interest than others. For instance, 
the model on square lattice ($t'=0$) has been studied in great detail near half filling.
Very few studies have explored the possibility for non-collinear and non-coplanar phases \cite{Chubukov1995, Ojeda1999}.
The half-filled case on triangular lattice has also been explored in search of spin liquid phases, as the large $U$ limit of the Hubbard model
leads to the spin-$1/2$ Heisenberg model which is frustrated on the triangular geometry.
There have been studies for the quarter filled case, which describes the physics of organic charge transfer salts \cite{Hotta2012, Amaricci2010, Tocchio2014}.
The filling fraction of $n=2/3$ has also been of interest due to its relevance to layered AgNiO$_2$ and NaCoO$_2$.
A systematic search for non-collinear and, in particular, non-coplanar phases over the full range of electronic densities has not been reported even at the mean field level.

\section{variational Mean-Field Scheme}
We determine the ground state phase diagrams by performing a systematic search in the phase space of various
ordered magnetic configurations. This is achieved by making use of the 
rotational invariance of the Hubbard Hamiltonian \cite{Schulz1990}. Before describing the details of the rotationally invariant scheme,
let us recall the commonly used unrestricted Hartree Fock (UHF) method \cite{Xu2011}. In this method, the interaction term is decoupled into
charge and spin sectors. Ignoring the 2nd order term in fluctuations, the 
Hamiltonian reduces to the well known form,

\begin{figure}
 \includegraphics[width=.9\columnwidth,angle=0, clip = 'True' ]{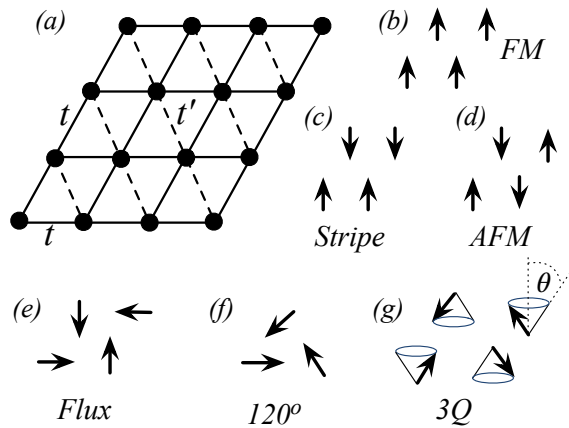}
 \caption{$($a$)$ A schematic view of the anisotropic triangular lattice with hopping parameters $t$ and $t'$ shown via solid and dashed lines, respectively.
 ($b$)-($g$) Building blocks of various long-range ordered magnetic phases. The cone angle $\theta$ connects the flux state to the AFM state via the 3Q state. 
}
\label{fig1}
\end{figure}    

\begin{eqnarray}
   H&=&-t(t')\sum\limits_{\langle ij \rangle
\sigma}c_{i\sigma}^{\dagger}c_{j\sigma} - \mu \sum_i n_i \nonumber  \\
   && + (1-\lambda) U\sum\limits_{i} [\langle n_{i\downarrow}\rangle
n_{i\uparrow} +\langle n_{i\uparrow}\rangle n_{i\downarrow}-
   \langle n_{i\downarrow} \rangle \langle n_{i\uparrow}\rangle ] \nonumber  \\
   &&- \lambda U \sum\limits_{i} [\langle s^{+}_{i} \rangle s^{-}_i+
   \langle s^{-}_{i} \rangle s^{+}_i-\langle s^{+}_{i}\rangle \langle s^{-}_{i}
\rangle],
\end{eqnarray}

\noindent
where, $s^{+}_{i}= c^{\dagger}_{i\uparrow}c^{}_{i\downarrow}$ and $s^{-}_{i} =
c^{\dagger}_{i\downarrow}c^{}_{i\uparrow}$ are the spin operators.
The variational parameter $\lambda$ determines the relative contribution of the
Hartree and the Fock terms in the mean-field decoupling.    

In the UHF approach, the site dependence of the quantum averages is
retained. Starting from a random guess for 
the various mean-field parameters, the Hamiltonian is diagonalized iteratively
until self-consistency is achieved. 
In principle, the self-consistent solution depends on the starting configuration
of mean-field parameters. Therefore, it is required to
use a large number of starting configurations. 
The reliability of the final solution obtained in this way depends to a large extent on the
complexity of the energy landscape. 
In addition,
$\lambda$ should be determined via an energy minimization over the various
self-consistent solutions.
Although, this scheme can capture the conventional ordered magnetic phases very
well, it can easily miss out on non-collinear and non-coplanar magnetic orderings.

Motivated by the appearance of such unusual magnetically ordered phases in a
Kondo lattice model, we search for similar magnetic phases
in the Hubbard model. For this purpose, we recast the mean-field decouplings in
a way that allows for a systematical search of 
a variety of ordered phases.
We begin by re-writing the Hamiltonian in a reference frame with
site-dependent spin-quantization axes \cite{Schulz1990}. This is done by performing local
SU(2) rotations of the quantization axes, given by,

\begin{eqnarray}
\displaystyle
\left[
\begin{array}{c}
 c_{i\uparrow}\\
 c_{i\downarrow}\\
\end{array} \right]
& = & \left[ \begin{array}{cc}
 cos(\frac{\theta_{i}}{2}) e^{i\frac{\phi_{i}}{2}} &  -sin(\frac{\theta_{i}}{2})
e^{i\frac{\phi_{i}}{2}}\\
 sin(\frac{\theta_{i}}{2}) e^{-i\frac{\phi_{i}}{2}} & cos(\frac{\theta_{i}}{2})
e^{-i\frac{\phi_{i}}{2}}\\
 \end{array} \right]
\left[
\begin{array}{c}
d_{ip}\\
d_{ia}\\
\end{array} \right] \nonumber \\
\vspace{0.5cm}
 & \equiv & {\cal R}(\theta_{i}, \phi_{i}) \left[
\begin{array}{c}
d_{ip}\\
d_{ia}\\
\end{array} \right],
\end{eqnarray}

\noindent
where $d_{ip}~(d_{ia})$ are annihilation operators for electron at site $i$ with
spin parallel (antiparallel) to the quantization axis.
A simple Hartree decoupling of the transformed Hamiltonian leads to,
\begin{eqnarray}
 H &=& -t(t')\sum\limits_{\langle ij \rangle} \sum\limits_{\sigma \sigma'}
(f_{\sigma \sigma'}d^{\dag}_{i\sigma}d_{j\sigma'}+ H.c)  \nonumber \\
      & & +U \sum\limits_{i} [\langle n_{ip} \rangle n_{ia} + \langle n_{ia}
\rangle n_{ip}- \langle n_{ip} \rangle \langle n_{ia} \rangle ], 
\end{eqnarray}
\vspace{0.5cm}

\noindent
where, $ n_{i\sigma} = d^{\dag}_{i\sigma}d_{i\sigma}$ and $\sigma$ can take two
values. 
The coefficients $f_{\sigma \sigma'}$ are explicitly given by,
\vspace{0.5cm}

\begin{eqnarray}
\displaystyle
\left[
\begin{array}{cc}
 f_{pp} & f_{pa} \\
 f_{ap} & f_{aa} \\
\end{array}
\right]
= {\cal R}^{\dag}(\theta_{i}, \phi_{i}). {\cal R}(\theta_{j}, \phi_{j})
\end{eqnarray}

\noindent
We consider two broad categories of variational states. First one is the set of general non-coplanar spiral states which
can be parameterized by a cone angle $\Theta$ and a spiral wave vector ${\bf q}$ \cite{Kumar2010spin}. The polar and azimuthal angles of the
spins are then given by, $\theta_i=\Theta$ and $\phi_i= {\bf q} \cdot {\bf r}_i$. 

The other set of states are the block-periodic states, where the spin configuration of a $m_x \times m_y$ block is repeated periodically to
generate the configuration over the entire lattice. The choice of these variational states is motivated by the presence of various
long-range ordered phases in the corresponding Kondo-lattice model. The flux state and the non-coplanar states shown in Fig. \ref{fig1}
are examples of the $2 \times 2$ block-periodic states. In the flux state the azimuthal angles of the neighboring spins differ by $\pi/2$
while the
polar angle remains equal to $\pi/2$. The non-coplanar state can be obtained from the flux state by changing the polar angles of the 
four spins as shown in Fig. \ref{fig1}. 
Similarly we take a $3 \times 3$ block to capture the \YK state, which can be connected to a ferromagnetic state by varying the polar angle from $\pi/2$ to 0.
In order to allow for the six-sublattice non-collinear charge ordered (NC-CO) state at $n=2/3$, we take a 
$6 \times 6$ cluster as the building block \cite{Reja2015}. The calculations have been performed in both canonical and 
grand canonical approaches. In the grand canonical (canonical) approach, we seek a self-consistent solution for a given fixed value of chemical potential (density) 
for all the variational states discussed above. In cases where more than one self-consistent solutions exists, the lower energy solution is selected and the
corresponding state is taken as the mean-field ground state at that value of the chemical potential (density). For the grand canonical approach, the density corresponding
to the ground state solution is calculated at the given value of the chemical potential and the phase diagrams are presented with density as a parameter.

\section{Results and Discussions}

In this section we present the results obtained by using the mean-field decoupling scheme described in the previous section.
For most of the calculations we used $N=64^2$ $k$-points in the first Brillouin zone. Results have also been checked for $N=128^2$ in some cases.
In the following subsection we discuss the $n$-$U$ phase diagrams for different values of $t'/t$.

\subsection{$n$-$U$ phase diagrams}

The $n$-$U$ phase diagrams are obtained within the grand canonical approach. The energy minimization over the 
two sets of variational states is performed for a fixed value of chemical potential $\mu$. The chemical potential is then systematically 
varied in order to obtain different average electronic densities. The total energy is computed at zero temperature, where the Fermi function simply becomes a step function.
The grand canonical approach has an advantage that it allows for the phase separation regions, which commonly arise in electronic systems, to be easily captured.

\begin{figure}[t!]
\includegraphics[width=.82\columnwidth,angle=0, clip = 'True' ]{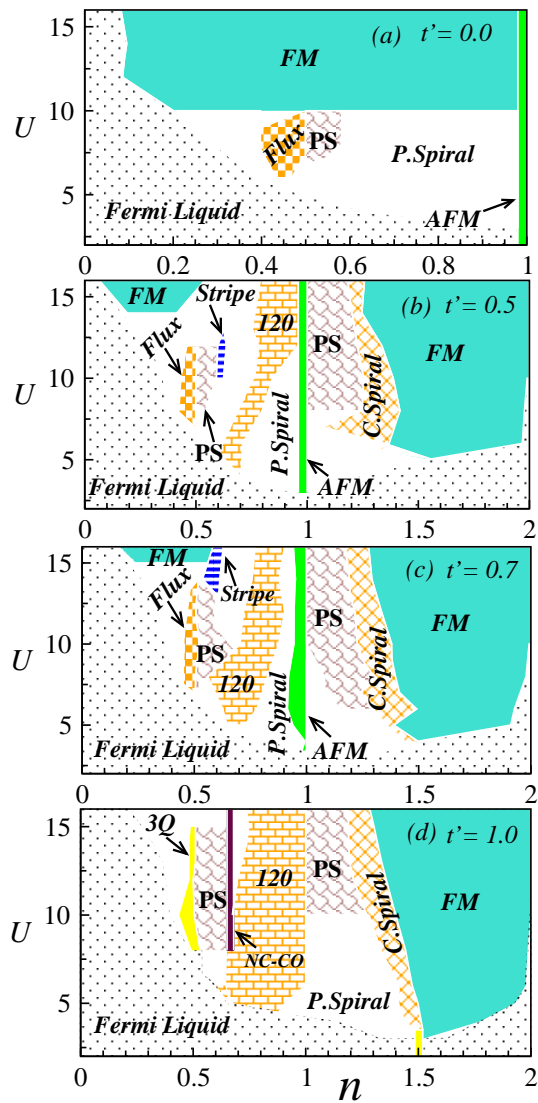}
 \caption{(Color online) $n$-$U$ phase diagrams obtained within grand canonical approach for,
($a$) $t'/t=0$,
  ($b$) $t'/t=0.5$, ($c$) $t'/t=0.7$ and ($d$) $t'/t=1$. PS denotes phase separation, P. spiral (C. spiral) refers to planar- (conical-) spiral 
phase, 3Q and NC-CO are the 4-sublattice and 6-sublattice ordered phases found at $n=1/2$ and $n=2/3$, respectively. 
}
\label{fig2}
\end{figure}

For $t'/t = 0$, the hopping connectivity is that of a square lattice. Therefore, in this case we display the density range $0 < n < 1$ in the phase diagram as the
system is particle-hole symmetric (see Fig. \ref{fig2} ($a$)). The square lattice phase diagram is dominated by three simple phases: a Fermi liquid state for weak to intermediate
$U$ and away from $n=1$, an antiferromagnetic (AFM) state near $n=1$ and a ferromagnetic (FM) state for large $U$ and away from $n=1$. In addition, in the intermediate $U$ range we find a Flux phase
and a narrow window of phase separation close to $n=1/2$. The flux phase is known to exits in the Kondo-lattice model with a classical approximation for local moments
\cite{Agterberg2000,Daghofer2005,Chen2010,Venderbos2012}. 
It has also been reported recently in he UHF study of a five-orbital Hubbard model for iron pnictides \cite{Luo2014}.
The planar spiral states are stable over a wide range of parameters.
Next, we present the phase diagram for the anisotropic case where the particle-hole symmetry does not hold. Therefore, the density range displayed in Fig. \ref{fig2} ($b$)-($d$) 
is $0<n<2$. 
The FM state is strongly suppressed by the triangular anisotropy in the low-density regime, whereas it still dominates the large density regime. This can be explained
by invoking the Stoner picture for ferromagnetism: the triangular lattice density of states is higher if the Fermi level is located close to $n=3/2$ and therefore the Stoner
criterion for ferromagnetism can be satisfied for relatively small values of $U$.
The flux state remains stable in a narrow window around intermediate $U$ and $n=1/2$. A new magnetically ordered state enters the phase diagram near $n=1$. This is the well known
$120^{\circ}$-state which is the classical ground state of a triangular lattice Heisenberg model. A wider window of phase separation appears in $n>1$ 
regime \cite{Poilblanc1989,LopezSancho1992,Igoshev2010}, where a 
conical spiral state also becomes stable. The phase diagram remains qualitatively similar between $t'/t = 0.5$ and $t'/t = 0.7$. However, the isotropic triangular
case ($t' = t$) gives rise to new and interesting phases. The first of these new phases is the 4-sublattice non-coplanar order near $n=1/2$. This unusual magnetic state, also known 
as 3Q state,
is of wide interest as it supports a non-vanishing scalar spin chirality and an associated quantized Hall conductivity \cite{Martin2008,Akagi2010a}. 
The quantized Hall conductivity is related to the 
nontrivial topology of the electronic band structure that the 3Q state induces in the Kondo Hamiltonians \cite{Kourtis2012}. 
Interestingly, the 3Q state was predicted to be stable in the Hubbard model at $n=3/2$ 
based on Fermi surface nesting arguments \cite{Martin2008}. 
We find that the 3Q state is stable at $n=3/2$ only for $U \leq 3.5$ and loses to FM order for larger values of $U$.
The second new state that appears exactly at $n=2/3$ is the recently reported $6$-sublattice 
non-collinear charge ordered (NC-CO) state. In addition, the $(\pi,\pi)$ ordered AFM state completely disappears, and instead the $120^{\circ}$-state
dominates the phase diagram near $n=1$, as expected \cite{Krishnamurthy1990,Jayaprakash1991}.

Having discussed the overall structure of the $n$-$U$ phase diagrams, we now examine more closely the effect of anisotropic hopping connectivity
on different magnetic phases at specific filling fractions.

\subsection{ $t'$-$U$ phase diagrams at $n=1/2$, $2/3$, $1$, $3/2$}

\begin{figure}
 \includegraphics[width=.82\columnwidth,angle=0, clip = 'True' ]{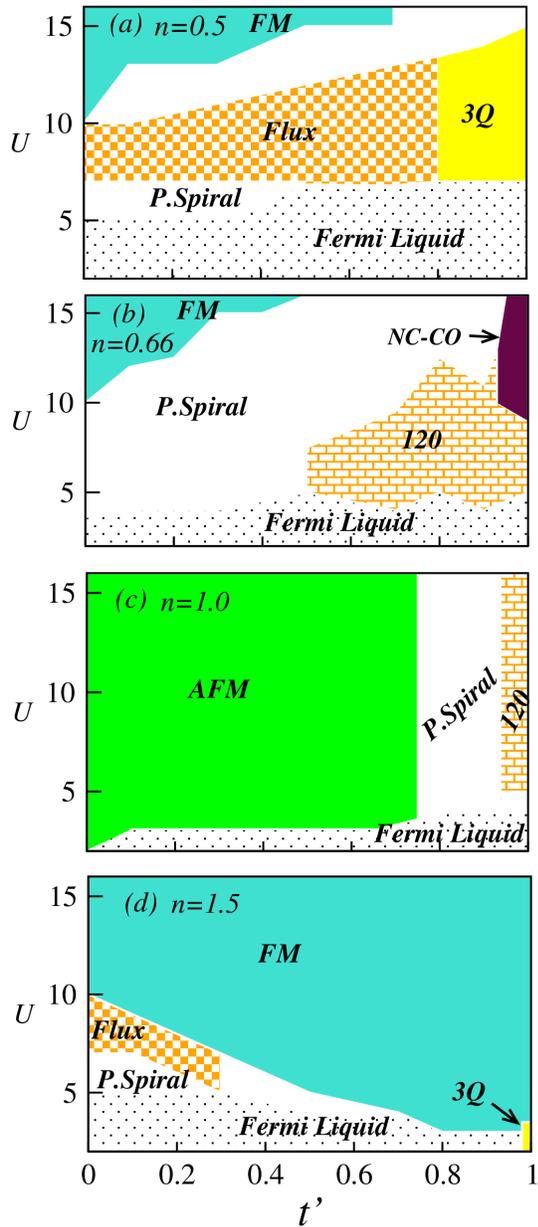}
\caption{(Color online) $t'$-$U$ phase diagrams obtained within canonical ensemble approach at commensurate values of average
electronic fillings, ($a$) $n=1/2$, ($b$) $n=2/3$, ($c$) $n=1$, and ($d$) $n=3/2$. The notation for different phases is same as that in Fig. \ref{fig2}.
}
\label{fig3}
\end{figure}

\begin{figure}
 \includegraphics[width=.96\columnwidth,angle=0, clip = 'True' ]{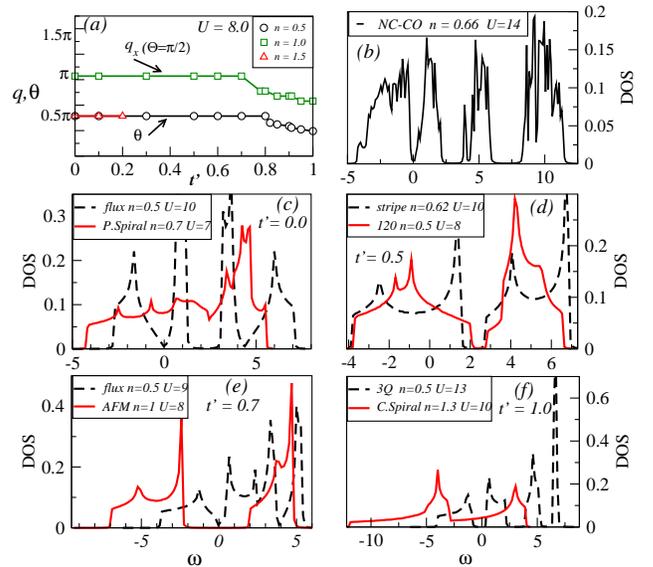}
\caption{(Color online) ($a$) Change in the variational parameters corresponding to the ground states for $n=1/2$, $n=1$ and $n=3/2$ as a function of
$t'$. The parameters are, the spiral wavevector $q_x = q_y$ for $n=1$, and the polar angle $\theta$ describing the $2 \times 2$ block-periodic states for $n=1/2, 3/2$.
For $n=3/2$ the ground state is FM for $t'/t > 0.2$.
($b$)-($f$) 
The electronic density of states for various ground states for different values of $t'$: ($b$) the NC-CO state at $n=2/3$ and $t'=1$, 
($c$) the flux and planar spiral states at $n=0.5$ and $n=0.7$, respectively, for $t'=0$
($d$) the stripe and \YK states at $n=0.62$ and $n=0.5$, respectively, for $t'/t = 0.5$ ($e$) the flux and the staggered AFM states for $t'/t=0.7$, and ($d$) the
non-coplanar 3Q and conical spiral states for $t'/t=1$.
} 
\label{fig4}
\end{figure}

In this subsection we describe the $t'$-$U$ phase diagrams obtained within canonical ensemble approach, where the minimization is performed over the variational parameters
for a fixed value of the average electronic density $n$.
We discuss the evolution of various phases as the lattice connectivity changes from square type to triangular type.
The non-collinear flux phase is lowest energy state for unfrustrated square lattice ($t'/t=0$) at 
$n=1/2$ for intermediate coupling strength. This state remains stable in a wide parameter regime,
$0 \leq t'/t \leq 0.8$. In fact, the stability window of the flux state widens in terms of the $U$ values upon increasing $t'/t$ (see Fig. \ref{fig3}($a$)).
For $t'/t > 0.8$, a non-coplanar state becomes stable which evolves into the 3Q state at $t'/t = 1$. 
The evolution from the flux state to the 3Q non-coplanar state is depicted by the change in variational angle $\theta$ ( see Fig. \ref{fig4} ($a$)).
For the flux state $\theta = \pi/2$, whereas for the perfect 3Q state $\theta \approx 0.98$.
The other effect of the triangular geometry for $n=1/2$ is to suppress the FM phase. The FM phase give way to planar spiral phases even for small values 
of the parameter $t'/t$. The transition from a non-magnetic Fermi liquid state to a magnetic state also occurs at a slightly larger value of $U$ for the triangular
lattice in comparison to the square lattice.

Another unusual six-sublattice magnetic order was recently reported in the Kondo-lattice model for average density of two electrons per three sites \cite{Reja2015}. 
In order to explore the possibility of this phase in the Hubbard model, we focus on the filling fraction of $n=2/3$ (see Fig. \ref{fig3}($b$)). We again find that the FM state is suppressed in favor of the planar spiral
states. Additionally the $120^{\circ}$ order is stable for $t'/t > 0.5$. It is interesting to note that the stability window of the $120^{\circ}$ state at $n=2/3$ is
even wider than that at $n=1$. Importantly, we find that the six-sublattice NC-CO order is the ground state
for $t'/t > 0.93$ and for $U > 7$. 

The half-filling, $n=1$, is the most commonly studied filling fraction in the Hubbard models on triangular and square lattice \cite{Laubach2015}. Within the mean-field, we find that the
AFM order is robust and is stable in the regime $0< t'/t < 0.7$. Beyond $t'/t = 0.7$, the AFM order continuously evolves towards the $120^{\circ}$ state. In Fig. \ref{fig4}($a$), 
we show the change in the spiral wave-vector $(q_x, q_y)$ from $(\pi,\pi)$ to $(2\pi/3, 2\pi/3)$. 
The half-filling case is also of interest for the possible existence of a spin-liquid state \cite{Tocchio2013}. 
However, our mean-field method does not capture the spin-liquid states and therefore we cannot comment on this competition in the present paper.

Finally we discuss the filling fraction $n=3/2$, which is interesting 
due to the presence of Fermi-surface nesting feature in the non-interacting Hamiltonian. In fact, the non-coplanar 3Q state was first predicted based on the nesting property
of the Fermi surface. Clearly, for small $U$ the 3Q state is the ground state due to the presence of a peculiar Fermi surface nesting in the non-interacting Hamiltonian \cite{Martin2008}.
Interestingly, the 3Q order loses to FM order for $U>3.5$. 
The relative stability of the 3Q state at $n=1/2$ and $n=3/2$ was also explored in the 
Kondo-lattice Hamiltonian with a conclusion that the window of stability of the 3Q state is wider for $n=1/2$ compared to $n=3/2$ \cite{Akagi2010a}.
Our calculations on the Hubbard model support the results of the corresponding Kondo-lattice model.
In sharp contrast to the quarter filling ($n=1/2$), the three-quarter filling favors FM state for increasing $t'/t$. As mentioned earlier, this can be
understood from the particle-hole asymmetry and large DOS in the $n>1$ regime for the triangular lattice.

The electronic spectrum corresponding to different phases is displayed in the plots for the density of states in Fig. \ref{fig4}($b$)-($f$). 
The DOS is defined as,
\begin{eqnarray}
D(\omega) &=& \frac{1}{N} \sum_{k} \delta(\omega - \epsilon_k) \approx \frac{1}{N} \sum_k \frac{\gamma/\pi}{[\gamma^2 + (\omega - \epsilon_k)^2]},
\end{eqnarray}

\noindent
where $\epsilon_k$ are the eigenvalues corresponding to the lowest-energy self-consistent solution, $\delta$ is the Dirac Delta function and $\gamma$ is the broadening parameter
which we take to be $0.02 t$ for calculations.
All the three possible varieties of the spectra are present: The spiral states are metallic, the 3Q non-coplanar, the $(\pi,\pi)$
AFM and the NC-CO states are insulating, and the flux state is semimettalic with a graphene-like DOS. A robust gap in the DOS is suggestive of the 
stability of these phases against higher-order quantum corrections.

\section{Effect of Inter-site Coulomb repulsion}
In this section we analyze the influence of nearest neighbor Coulomb repulsion on various magnetic phases discussed so far.
We begin by extending the Hubbard model Eq. (1) by including a nn repulsion. The extended Hubbard Hamiltonian is given by,

\begin{eqnarray}
H' &=& H +  V (V') \sum_{\langle ij \rangle} n_{i} n_{j}.
\end{eqnarray}

\noindent
The parameters $V$ and $V'$ denote the strength of the Coulomb repulsion along inequivalent directions on the anisotropic triangular lattice.
It is well known that nn Coulomb repulsion in bipartite lattices favors charge ordering \cite{Amaricci2010}. However, the triangular geometry frustrates the
checkerboard type charge ordering and leads to exotic charge ordered phases, such as the pinball liquid and inverse pinball liquid \cite{Miyazaki2009,Merino2013,Watanabe2008,Watanabe2005}.
Here our primary focus is to study the effect of nn Coulomb repulsions on the various magnetic phases found in the Hubbard model.

\begin{figure}
 \includegraphics[width=.96\columnwidth,angle=0, clip = 'True' ]{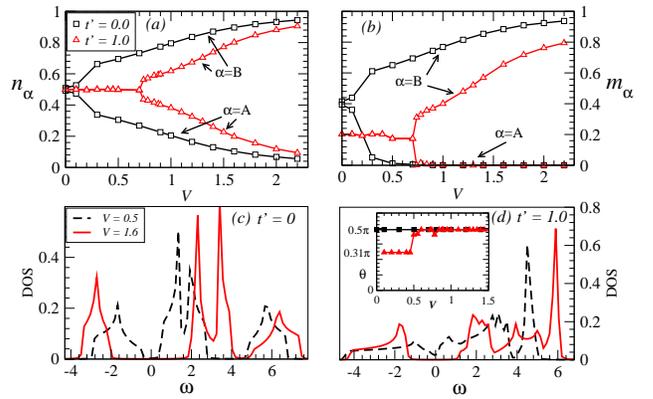}
\caption{(Color online) Results on the effect of Coulomb repulsion at $n=1/2$. The value of, ($a$) the local charge densities, and ($b$) the local magnetic moments 
at two inequivalent sites as a function of $V$. The inset in ($d$) shows the change in variational angle $\theta$ which connects the flux state ($\theta = \pi/2$)
to the 3Q state ($\theta \sim 0.31 \pi$). ($c$)-($d$) Density of states of the spin-charge ordered phases at different values of $V$ for ($c$) $t' = 0$ and ($d$) $t'=1$. 
These results are obtained for $U=8$ and $V'=0$.
}
\label{fig5}
\end{figure}

We present results for the quarter-filled case where the interesting mean-field ground states are the flux state, for $t'/t \leq 0.8$, and non-coplanar states
for $t'/t > 0.8$. We use the variational scheme corresponding to the block-periodic states discussed in Sec. III. This allows for four inequivalent sites in terms of
the spin and charge structures. The planar structure is fixed to be of the flux type and the polar angle is varied to obtain magnetic phases that interpolate between 
the flux and the ($\pi,\pi$) AFM state via the 3Q state (see Fig. \ref{fig1}).
Results for $V'=0$ are summarized in Fig. \ref{fig5}. We find that the flux state ($t'/t = 0$) is compatible with stripe charge ordering. Indeed, the charge disproportionation
$n_B - n_A$ becomes finite as soon as $V \neq 0$ (see Fig. \ref{fig5} ($a$)). The magnetic structure retains the flux character, however, the magnetic moments become unequal on
sites $A$ and $B$ (see Fig. \ref{fig7} ($a$)-($c$) ). For very large values of $V$ the magnetic moment becomes vanishingly small at one of the sites, leading to
a strongly charge ordered AFM state. In fact, a similar state has been reported in the quarter-filled Hubbard-Holstein, and Kondo lattice models on square lattice \cite{Kumar2008charge,Misawa2013}.
The nature of the magnetic order is depicted by the value of the variational polar angle $\theta$ corresponding to the minimum energy
self-consistent solution. $\theta = \pi/2$ remains constant for $t'/t = 0$ (see inset in Fig \ref{fig5} ($d$)).

In contrast to the flux state, the 3Q state competes with charge ordering. It is easy to understand this difference from the energetics: charge ordering leads to lowering of energy
via opening an energy gap in the spectrum. Since the flux state is gapless, a gap opening lowers the energy and therefore the ground state develops a charge ordering
even for small values of $V$. However, the 3Q state is already gapped with the origin of the gap tied to peculiar magnetic structure. Therefore, in order to further lower the
energy the charge ordering must open a gap that off-sets the 3Q gap.
Starting with $t'/t = 1$, the charge disproportionation remains zero until $V = 0.7$.
A weak reduction of the magnetic moment occurs near $V=0.5$ together with a deviation from the ideal 3Q structure. Near $t'/t = 0.7$, $n_B - n_A$ becomes finite, and the 
magnetic moment at site $A$ vanishes, leading to the same charge ordered AFM state as for $t'=0$. Therefore, in this case the 3Q state is destabilized at the 
onset of charge ordering. The density of states (fig. \ref{fig5}($c$)-($d$)) show that a gap opens in the electronic spectrum owing to the charge ordering. Thus, the semimettalic flux state is turned
into a fully gapped flux state, and eventually into a gapped collinear AFM for large $V$.
The results for the isotropic repulsive ($V' = V$) case are qualitatively similar to those discussed above. The main difference is that the
charge disproportionation is smaller for $V' = V$ compared to $V' = 0$.

\begin{figure}
\includegraphics[width=.96\columnwidth,angle=0, clip = 'True' ]{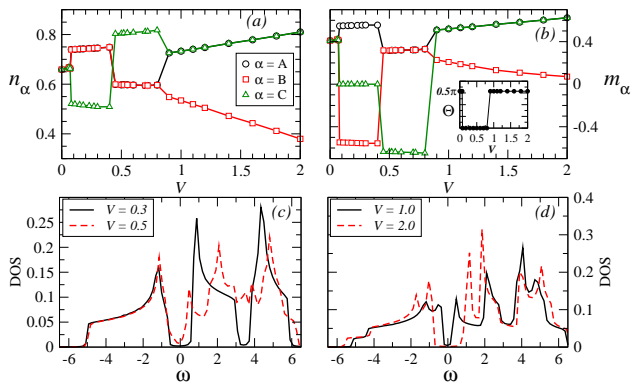}
\caption{(Color online) Results on the effect of Coulomb repulsion at $n=2/3$. The value of, ($a$) the local charge densities, and ($b$) the local magnetic moments 
of three inequivalent sites as a function of $V$. 
The inset in ($b$) shows the change in variational angle $\Theta$ which connects the FM state ($\Theta = 0$)
to the \YK state ($\Theta = \pi/2$). Density of states of the spin-charge ordered phases at different values of $V$. These results are obtained for $U=8$ and $V'=0$.
}

\label{fig6}
\end{figure}

Finally, we study the effect of Coulomb repulsion on the \YK phase. The results are presented in Fig. \ref{fig6}.
The \YK state is destabilized in favor of a collinear state containing three inequivalent sites. Two of the sites have equal charge 
density which is larger than that on the third ($n_A = n_B > n_C$). The
magnetic moments on the low-charge sites vanish, while those on the other two sites are antialligned (see Fig. \ref{fig6} ($a$)-($b$) and \ref{fig7}($d$)).
The resulting spin-charge order can be visualized as two interpenetrating lattices: an AFM ordered honeycomb lattice and a nonmagnetic triangular lattice.
This spontaneous separation into two sublattices is similar to what happens in the pinball-liquid phase. The crucial difference being the insulating character of the
DOS in the present case.
For $V > 0.4$, the charge densities on the three sites are
given by $n_A = n_B < n_C$, and all three sites have finite magnetic moments. The resulting state with up-up-down structure have also been 
found as the ground state of the Ising spin Kondo lattice model.
For $V \geq 0.9 $ the charge densities at three sites are given by $n_A = n_C > n_B$, and the 
magnetic state again becomes non-collinear (see inset in Fig. \ref{fig6}($b$)). Upon a further increase of $V$ the
charge density as well as the magnetic moment on one of the sites keep reducing, giving rise to an effective honeycomb lattice for the electrons in the 
large $V$ limit. 

\begin{figure}
\includegraphics[width=.90\columnwidth,angle=0, clip = 'True' ]{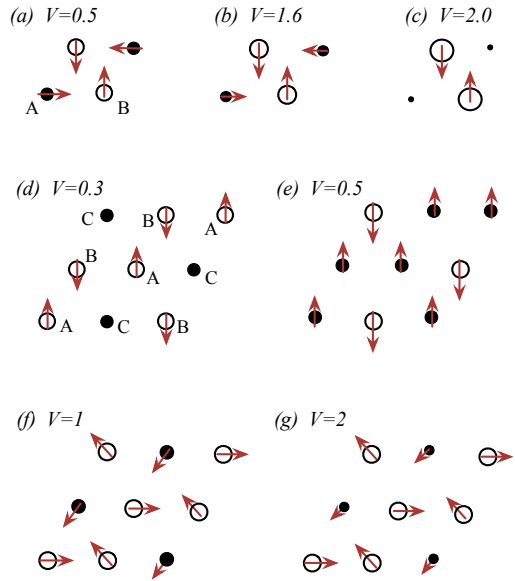}
\caption{(Color online) 
A schematic view of the spin-charge ordered phases that appear for finite inter-site Coulomb interaction for filling fraction, ($a$)-($c$) $n=1/2$, $t'=0$ 
and ($d$)-($g$) $n=2/3$, $t'=t$.
The size of the circles represent the local charge density. The higher density sites are shown as open circles for clarity.
}
\label{fig7}
\end{figure}

\section{Conclusions}

A systematic search is carried out for non-collinear and non-coplanar magnetic ground states in the Hubbard model. 
Using a mean-field decoupling scheme that 
treats collinear, non-collinear and non-coplanar order on equal footing, we uncover a rich phase diagram for the Hamiltonian on anisotropic triangular lattice.
The most notable of the unconventional magnetic ground states are, (i) the four-sublattice flux order, (ii) the 3Q non-coplanar state, and (iii) the six-sublattice non-collinear charge ordered state.
These states have been found as ground states of the Kondo lattice model on triangular lattice in previous studies \cite{Martin2008,Akagi2010a,Kumar2010,Reja2015}.
The effect of nearest neighbor Coulomb interactions on these magnetic states is also investigated, and leads to some fascinating spin-charge ordered phases.
The flux state is found to be compatible with a stripe-like charge ordered arrangement, whereas, the non-coplanar 3Q state competes with charge ordering and becomes unstable
upon increasing the Coulomb interaction strength beyond a critical value. The spin-charge ordered state found at $n=1/2$ is a triangular lattice version of the
state found in quarter-filled Hubbard-Holstein and Kondo lattice models \cite{Kumar2008charge,Misawa2013}.
The \YK state gives way to a sequence of spin-charge ordered states which lead to spontaneous decoupling of the triangular lattice into a honeycomb lattice and a triangular lattice.
For a certain range of parameter values, one of the sublattice is magnetic while the other is nonmagnetic. This highly resembles the pinball-liquid state reported 
in theoretical studies of multiband extended Hubbard model on triangular lattice \cite{Hotta2006,Hotta2006a,Miyazaki2009,Ralko2015}. 
The phase diagrams presented here set the stage for further studies to analyze the stability of these phases. A number of states reported here possess a robust gap in the 
electronic spectrum, and therefore are likely to remain stable against higher order quantum effects.
Given the elementary status of the Hubbard model and the recent interest in unconventional magnetic ordering, our results should be of general interest. 
In particular, the reported phases are
of relevance to various triangular lattice systems, such as, the layered AgNiO$_2$ \cite{Wawrzynska2008,Wawrzynska2007a}, 
the NaCoO$_2$ \cite{Mukhamedshin2014,Mukhamedshin2014a,Mukhamedshin2004,Bernhard2004,Alloul2009} and the organic charge-transfer salts \cite{Hotta2012,Kagawa2013}. 
Many of the phases presented in this work are similar to the spin-charge ordered phases reported experimentally in these materials. While material-specific models have been proposed
to understand these phases, it is interesting to see that the most elementary model for magnetism -- the single-band Hubbard model -- supports many of the unconventional spin-charge ordered
states.

\section{ACKNOWLEDGMENTS}

The calculations were performed using the High Performance Computing Facility at
IISER Mohali. 
K.P. acknowledges support via CSIR/UGC fellowship.
S.K. acknowledges support from DST, India.

\bibliographystyle{apsrev4-1} 

%


\end{document}